\documentclass[aps,prb,twocolumn,superscriptaddress,longbibliography]{revtex4-1}
\usepackage[version=3]{mhchem} 

\usepackage{dcolumn}
\usepackage{hyperref}%
\usepackage{graphicx}
\usepackage{array}
\usepackage{dcolumn}
\usepackage{bm}
\usepackage{float}
\usepackage{enumitem}
\usepackage[colorinlistoftodos, textwidth=4cm, shadow]{todonotes}
\newcommand{\degrees}{$^{\circ}$}

\newcommand{\FGT}{Fe$_{5}$GeTe$_2$~}
\newcommand{\FGTx}{Fe$_{5-x}$GeTe$_2$~}

\newcommand{\FCGT}{Fe$_{5-y}$Co$_y$GeTe$_2$~}
\newcommand{\TC}{$T_{\rm C}$~}
\newcommand{\TN}{$T_{\rm N}$~}
\newcommand{\HperpC}{$H$ $\perp$ $c$~}
\newcommand{\HparC}{$H \parallel c$~}
\newcommand{\MT}{$M$($T$)~}
\newcommand{\MH}{$M$($H$)~}

\renewcommand{\theequation}{\textbf{\arabic{equation}}}
\renewcommand{\thefigure}{\textbf{\arabic{figure}}}
\renewcommand{\thetable}{\textbf{\arabic{table}}}

\begin{document}

%

\title{Tuning Magnetic Order in the van der {Waals} Metal Fe$_5$GeTe$_2$ by Cobalt Substitution}

\author{Andrew F. May}
\email{mayaf@ornl.gov}
\affiliation{Materials Science and Technology Division, Oak Ridge National Laboratory, Oak Ridge, Tennessee 37831, United States}

\author{Mao-Hua Du}
\affiliation{Materials Science and Technology Division, Oak Ridge National Laboratory, Oak Ridge, Tennessee 37831, United States}

\author{Valentino R. Cooper}
\affiliation{Materials Science and Technology Division, Oak Ridge National Laboratory, Oak Ridge, Tennessee 37831, United States}

\author{Michael A. McGuire}
\affiliation{Materials Science and Technology Division, Oak Ridge National Laboratory, Oak Ridge, Tennessee 37831, United States}

\date{\today}

\begin{abstract}
\textbf{Fe$_{5-x}$GeTe$_2$ is a van der Waals material with one of the highest reported bulk Curie temperatures, \TC $\approx$310\,K.  In this study, theoretical calculations and experiments are utilized to demonstrate that the magnetic ground state is highly sensitive to local atomic arrangements and the interlayer stacking.  Cobalt substitution is found to be an effective way to manipulate the magnetic properties while also increasing the ordering temperature.  In particular, cobalt substitution up to $\approx$ 30\% enhances \TC and changes the magnetic anisotropy, while $\approx$50\% cobalt substitution yields an antiferromagnetic state. Single crystal x-ray diffraction evidences a structural change upon increasing the cobalt concentration, with a rhombohedral cell observed in the parent material and a primitive cell observed for $\approx$46\% cobalt content relative to iron.  First principles calculations demonstrate that it is a combination of high cobalt content and the concomitant change to primitive layer stacking that produces antiferromagnetic order.  These results illustrate the sensitivity of magnetism in \FGTx to composition and structure, and emphasize the important role of local structural order/disorder and layer stacking in cleavable magnetic materials.}
\end{abstract}

\maketitle


\section{Introduction}

The disparity between interlayer and intralayer bonding forces in exfoliable materials typically results in weak interplane magnetic interactions.\cite{Bhimanapati2015}\cite{Bhimanapati2015,Park2016,Duong2017,McGuire-halides2017,Burch2018,Li2019}  This can lead to low magnetic ordering temperatures, particularly in insulators with long direct or multi-atom exchange paths between layers.  The case is more complicated in cleavable metals, where delocalization makes bonding and magnetic interactions more difficult to understand.  One of the highest reported Curie temperatures ($T_{\rm C}$) in a cleavable bulk material is found in metallic \FGTx with bulk $T_{\rm C}$ as high as 310\,K;\cite{Stahl2018,ACSNano,May2019} magnetic order was demonstrated to exist at similarly high temperature in exfoliated flakes with thicknesses approaching 10\,nm.\cite{ACSNano}  Magnetic order above 200\,K is also found in metallic Fe$_5$Ge$_2$Te$_2$\cite{Jothi2019} and Fe$_{3-x}$GeTe$_2$,\cite{Deiseroth2006,Chen2013,Verchenko2015,May2016} the latter of which has garnered significant attention with magnetic order demonstrated in the monolayer limit.\cite{Fei2018,Deng2018}  Such itinerant magnetism is strongly coupled to the electronic structure and atomic-scale disorder and site substitutions can have a significant impact on \TC and the magnetic anisotropy, as shown for Fe$_{3-x}$GeTe$_2$.\cite{Drachuck2018,Tian2019,Hwang2019}  The magnetism in these cleavable metals is expected to be quasi-2D, and the ferromagnetic (FM) order has been predicted to persist in monolayer Fe$_{5}$GeTe$_2$.\cite{Joe2019}  Relatively weak interplane interactions have indeed been demonstrated by inelastic neutron scattering for Fe$_{3-x}$GeTe$_2$.\cite{Calder2019}  This implies an ability to tune the interplane coupling and modify the magnetic response.  Such an effect has been observed in insulating CrI$_3$, where antiferromagnetic (AFM) or ferromagnetic coupling between layers depends on the cell symmetry (layer stacking).\cite{Sivadas2018,Jiang2019,Song2019,Li2019}  In RuCl$_3$, the N\'{e}el temperature \TN varies by $\approx$100\% with changes in layer stacking ($\approx$7 to 14\,K).\cite{Cao2016} 

\begin{figure}[ht!]%
\includegraphics[width=\columnwidth]{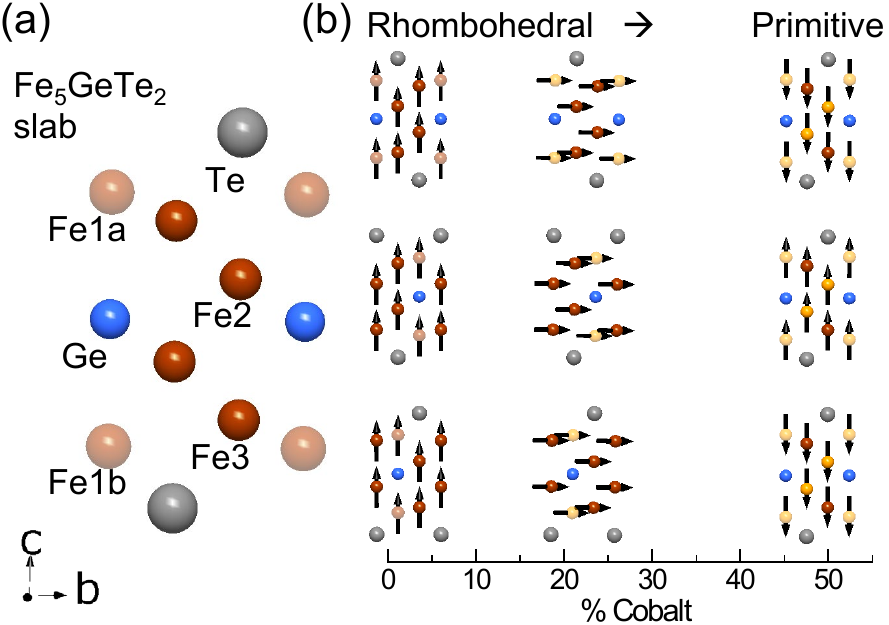}%
\caption{(a) Schematic of \FGT layer with atomic types and positions labeled. Fe1a,b represent a split site that allows for local atomic order/disorder, and the associated split site of Ge is not illustrated for simplicity. (b) Change in lattice type (rhombohedral/primitive) with cobalt doping (gold spheres) and corresponding evolution of magnetic order and anisotropy (black arrows). Preferred cobalt locations are shown for Fe$_4$CoGeTe$_2$ and Fe$_2$Co$_3$GeTe$_2$ based on first principles calculations; all Fe1 sites are filled for ease of viewing.  Rhombohedral centering with ABC layer stacking is shown for Fe$_5$GeTe$_2$ and Fe$_4$CoGeTe$_2$, while primitive centering with AAA layer stacking is shown for Fe$_2$Co$_3$GeTe$_2$.  In real crystals, the Fe$_4$CoGeTe$_2$ composition is expected to possess stacking disorder associated with the competition between these two stacking sequences.}%
\label{Structure}%
\end{figure}

Through a combined experimental and theoretical effort, this research establishes a connection between the magnetic ground state and the local atomic order and layer stacking in Fe$_{5}$GeTe$_2$, thereby demonstrating control over the room temperature magnetic properties. \FGTx contains local atomic order/disorder due to a split site (Fe1a,b in Fig\,\ref{Structure}(a)), and the associated magnetic sublattice orders near 120\,K despite \TC\,=\,310\,K.   The density functional theory (DFT) calculations reported here find that a distribution of Fe atoms on Fe1a and Fe1b sites yields a FM ground state, while the energetically unfavorable configuration of occupying only the Fe1a positions has an antiferromagnetic ground state with small Fe1 moments.   Experimentally, cobalt substitution up to $\approx$30\% is found to enhance \TC while changing the anisotropy from easy-axis [001] to easy-plane in these trigonal materials.  At higher cobalt concentrations ($\approx$45\%-55\%), the magnetization has a predominantly AFM character with \TN remaining between $\approx$290-340\,K.  X-ray single crystal diffraction evidences a change in layer stacking from rhombohedral to primitive for $\approx$50\% cobalt.  In turn, DFT calculations suggest that it is a combination of the primitive layer stacking and the high cobalt content that leads to an AFM ground state.  In total, these results demonstrate control of magnetic coupling, ordering temperature and anisotropy via chemical substitution, local atomic arrangements and layer stacking in a metallic vdW material.

\section{Methods}

Cobalt-containing crystals were grown using the same iodine-assisted approach as utilized for Fe$_{4.87(2)}$GeTe$_2$,\cite{ACSNano,May2019} using initial compositions of Fe$_{5-y}$Co$_y$GeTe$_2$. The raw elements were sealed in evacuated silica ampoules and heated to 750\degrees C at 120\degrees/h, followed by an isothermal step for 1-2 weeks at 750\degrees C.    The ampoules were quenched into ice-water and iodine was washed from the surfaces of crystals with solvents (ethanol and/or isopropanol with an acetone rinse). An attempt to form Co$_{5}$GeTe$_2$ in the presence of iodine yielded only binary compounds.

Single-crystal x-ray diffraction data were collected using a Bruker D8 Quest with a nitrogen cold stream at 220\,K by mounting the crystals in paratone oil.  Structural solution and refinement were performed using ShelX after data reduction via SMART-Plus.\cite{Sheldrick2015}  Small crystals ($<$100$\mu$m) were selected from batches where the larger  crystals ($>$1\,mm) had experimental compositions of 28 and 46\% cobalt. The experimental cobalt concentrations used throughout the paper were obtained via energy dispersive spectroscopy (EDS).  EDS data were collected using a Hitachi TM-3000 scanning electron microscope equipped with a Bruker Quantax 70 EDS detector system; the accuracy is expected to be 1-2 atomic \%. The cobalt concentration was determined as that relative to the total transition metal content and a net change in total transition metal content was not detected within the resolution of the instrument. X-ray diffraction data from the crystal facets were collected using a PANalytical X'Pert Pro MPD utilizing a Cu K$\alpha_1$ ($\lambda$=1.5406\,\AA) incident beam monochromator.  Le Bail fitting was then performed using the program FullProf.\cite{FullProf} Magnetization data were collected in SQUID magnetometers from Quantum Design (MPMS, MPMSXL, MPMS3), using the DC approach with data obtained upon cooling in an applied field.  Isothermal \MH data are plotted as a function of applied field and were collected upon decreasing the field toward zero; demagnetization effects do not impact the conclusions and are discussed in the Supplemental Materials.\cite{SupportingInfo}   

First principles calculations were based on density functional theory with Perdew-Burke-Ernzerhof  (PBE) exchange-correlation functional\cite{PerdewGGA1996} as implemented in the VASP code.\cite{Kresse1996a} The kinetic energy cutoff of the plane-wave basis is 310\,eV. The projector augmented wave method was used to describe the interaction between ions and electrons.\cite{Kresse1999} A 6$\times$6$\times$1 k-point mesh was used for the 2$\times$2$\times$1 of \FGT and the 2$\times$2$\times$2 cell of 20\% cobalt substitution (Fe$_4$CoGeTe$_2$).  A 6$\times$6$\times$2 k-point mesh was used for the 2$\times$2$\times$2 cell of 60\% cobalt substitution (Fe$_2$Co$_3$GeTe$_2$). A test calculation using a denser 8$\times$8$\times$2 k-point mesh for 60\% cobalt changed the energy difference between FM and AFM orderings by only 0.5\,meV/f.u.. The lattice parameters are fixed at the experimentally measured values obtained at $\approx$ 220\,K for 0 and 46\% cobalt.  The atomic positions are optimized until the force on each atom is less than 0.02 eV/\AA.  Appropriate supercells were taken to probe the energies of different atomic arrangements on the Fe1a / Fe1b split site as well as to explore antiferromagnetic coupling along the $c$-axis.

\section{Results and Discussion}

\subsection{Crystal Growth and Structural Information}

The structural unit of \FGT is shown in Fig.\ref{Structure}(a) and atomic-level disorder on the Fe1 sublattice likely impacts many physical properties. This sublattice is a split-site that resides either above (Fe1a) or below (Fe1b) the neighboring Ge atom due to bond distance restrictions.  The Ge atom also resides at a split site as it moves down (up) to adjust for occupation of Fe1a (Fe1b); the Ge split site is not illustrated in Fig.\ref{Structure} for simplicity. The Fe1 sublattice also hosts vacancies but the total Fe content has not been manipulated. Single crystals that we have grown in the presence of iodine using various nominal concentrations of Fe always possess the same composition, which was determined to be Fe$_{4.87(2)}$GeTe$_2$ by refinement of single crystal x-ray diffraction data and Fe$_{4.7(2)}$GeTe$_2$ by wavelength dispersive spectroscopy.\cite{ACSNano}

\begin{figure}[h!]%
\includegraphics[width=\columnwidth]{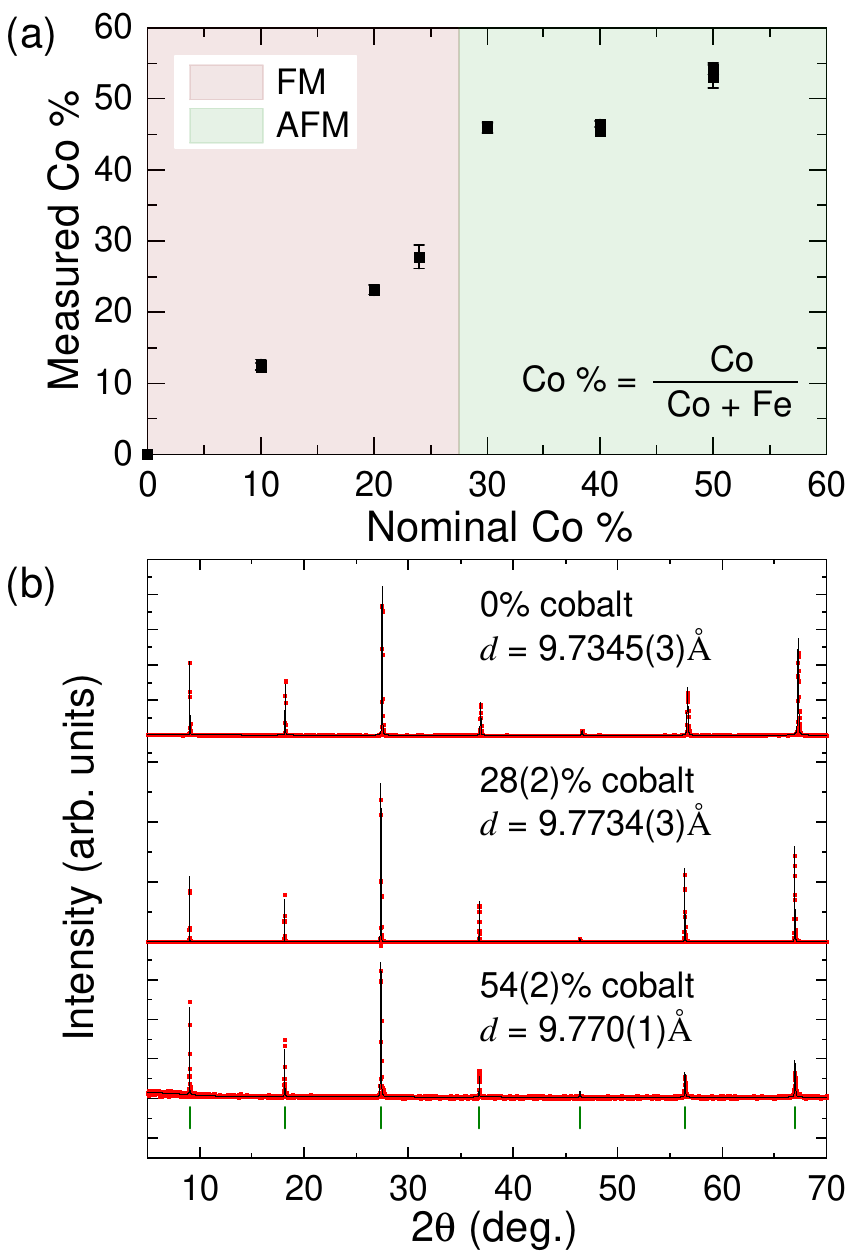}%
\caption{(a) Experimental versus nominal cobalt concentration with shaded regions where ferromagnetic (FM) and antiferromagnetic (AFM) behavior are observed.  (b) X-ray diffraction data from crystal facets;  these 00$l$ reflections (red) were fit (black lines) to obtain the layer spacings $d$ for select compositions.}%
\label{EDS}
\end{figure}

The unit cell is composed of three \FGTx layers with rhombohedral layer stacking,\cite{Stahl2018,ACSNano} as illustrated on the left of Fig.\,\ref{Structure}(b) (space group $R\bar{3}m$).  This ideal description is complicated by a phase transition near 570\,K that involves layer stacking and induces significant stacking disorder in samples that are cooled slowly to room temperature.\cite{May2019}  The primitive layer stacking shown at the right edge of Fig.\,\ref{Structure}(b) was speculated to be the most common stacking fault.\cite{Stahl2018} Interestingly, this is the structural model we report herein for a crystal with $\approx$46\% cobalt, as discussed below and summarized in Table \ref{Structure}.  The coupling and orientations of the magnetic moments are shown in Fig.\,\ref{Structure}(b) at representative compositions, and these trends with cobalt substitution are established via the data and calculations discussed below.

\begin{table*}
\caption{\textbf{Refined structural parameters for (Fe,Co)$_5$GeTe$_2$ from single-crystal x-ray diffraction data.} Space group $P\bar{3}m1$ (No. 1647); $a$ = 4.0196(2)\AA, $c$ = 9.8000(6)\AA, $T$ = 220\,K, $R1$  =  0.0466, $wR2$ = 0.129 and Goodness of Fit (GooF) = 1.36 for all 232 reflections after merging using 17 parameters and zero constraints. The refinement assumed a fixed composition of 50\% iron/cobalt and the expected composition is $\approx$46\% cobalt.}
\begin{tabular}[c]{|c|c|c|c|c|}
\hline
atom  &  $x,y,z$ & Wykoff position & occupancy &   \\
\hline
Te1 &  $\frac{1}{3}$,$\frac{2}{3}$,0.34347(11)  & 2d &  1 &      \\
Ge1 &  0,0,0.9823(7) & 2c & 0.5 & (split site)\\ 
Fe1/Co1 &  0,0,0.2318(4)   & 2c & 0.5 & (split site) \\ 
Fe2/Co2 &   $\frac{2}{3}$,$\frac{1}{3}$,0.1808(3)  & 2d & 1 &  \\
Fe3/Co3 &   $\frac{1}{3}$,$\frac{2}{3}$,0.0760(3)  & 2d & 1 & \\ 
\hline
 &  $U_{11}$ &$U_{22}$ & $U_{33}$ & $U_{12}$  \\
\hline
Te1 & 0.0074(4)& 0.0074(4) &0.0076(5)& 0.0037(2)  \\
Ge2 & 0.0029(6) &0.0029(6)& 0.013(5) &0.0015(3)\\
Fe1/Co1 & 0.0030(9)& 0.0030(9)& 0.0066(17)& 0.0015(5)\\  
Fe2/Co2 & 0.0118(6) &0.0118(6)& 0.0103(11) &0.0059(3) \\
Fe3/Co3 & 0.0058(6) &0.0058(6) &0.0053(10)& 0.0029(3) \\
\hline
\end{tabular}
\label{tab:refine2}
\end{table*}

We employed first principles calculations based on density functional theory to investigate the preferential occupation on the Fe1 sublattice and the substitution of cobalt into the lattice.  The lattice parameters were fixed and all atomic positions were relaxed locally in response to the different atomic orderings imposed for the Fe1 sublattice.  The calculations revealed a significantly lower energy when the Fe1 atoms are distributed on both Fe1a and Fe1b positions as opposed to occupation on only Fe1a.  The results are summarized in Table\,\ref{DFT}, where energies relative to the ground state are provided and results for cobalt substitution are also included but discussed later.  Distributing Fe atoms on both Fe1a and Fe1b in a checkerboard pattern is 123\,meV/f.u. lower in energy than only occupying the Fe1a position. A striped distribution of atoms on Fe1a and Fe1b is within 3\,meV of the checkerboard atomic order.  These results suggest that distributing Fe across Fe1a,b is most energetically favorable, but that significant disorder will likely exist in real crystals; this is consistent with previous STEM observations.\cite{ACSNano}  The atomic order with only Fe1a positions occupied has a non-centrosymmetric unit cell as in the model put forth by Stahl \textit{et al} with space group $R3m$.\cite{Stahl2018}  Distributing atoms on both Fe1a and Fe1b seems to be more energetically favorable because it reduces the local density and distributes iron atoms across the slab so that more favorable bond distances can be achieved throughout the entire \FGT structural motif.  Relaxed atomic positions can be viewed in the structural files uploaded as Supplemental Materials.

We grew single crystals using an iodine assisted reaction, and the crystals obtained from a given growth were found to have similar compositions and magnetic properties.  Individual crystals were selected and energy dispersive spectroscopy (EDS) was performed prior to characterization by diffraction off as-grown or cleaved facets, magnetization, specific heat or electrical resistivity measurements.  The electrical resistivity data demonstrate that all compositions are metallic with a large residual resistivity, and the specific heat data suggest a decreasing electronic contribution with increasing cobalt content (see Supplemental Materials). 

The relationship between nominal cobalt concentration and the experimentally determined concentration is shown in Fig.\,\ref{EDS}(a). A good correspondence between the experimental and nominal cobalt concentrations was observed up to $\approx$25\% cobalt.   Crystals containing $\approx$30-45\% cobalt were not successfully synthesized; this may point to a miscibility gap, but a detailed investigation was not pursued.   The concentration of cobalt provided is relative to the total transition metal content and the error bars are the standard deviations obtained by averaging many points collected at various spots on a given crystal.  

X-ray diffraction off the facets of as-grown crystals was utilized to assess the interlayer spacing at a few cobalt concentrations, with data shown in Fig.\,\ref{EDS}(b).  Le Bail fitting to these data reveals a small increase in the interlayer spacing $d$ as cobalt is incorporated into the lattice.    The diffraction patterns in Fig.\,\ref{EDS}(b) only contain 00$l$ reflections and thus do not probe the lattice centering directly.  Assuming trigonal symmetry, the data can be indexed using a rhombohedral cell were $c$=3$d$ ($l$=3$n$ reflection indexing) or a primitive unit cell with $d=c$ ($l$=$n$ indexing) and thus robust information about $d$ can be obtained.  Due to the introduction of strain and/or stacking disorder upon grinding for powder diffraction, single crystal diffraction experiments were utilized to probe the symmetry in select \FCGT crystals.

X-ray single-crystal diffraction data were collected to assess how cobalt substitution impacts the average symmetry.  Crystals were selected from growths that produced average cobalt concentrations of 28 and 46\%.   The diffraction data for the crystal from the 28\% cobalt growth were not suitable for structural solution due to stacking disorder (significant streaking along $l$ for $h0l$ reflections as shown in the Supplemental Materials).  For the higher cobalt content ($\approx$46\%), stacking faults were not an issue and a structural model was developed in space group $P\bar{3}m1$. The data were indexed to a primitive unit cell with $a$=4.0196(2)\AA\, and $c$=9.8000(6)\AA\, at 220\,K, and the refined atomic positions and thermal parameters are summarized in Table \ref{Structure}. A comparison to the structure of \FGTx reveals that cobalt substitution causes a slight contraction within the $ab$-plane and an increase in the slab thickness and interlayer spacing. These results are consistent with the increase in layer stacking observed by diffraction from the crystalline facets. We note that a twin law was required for the structure solution.  Given that cobalt substitution drives a stacking transition, it is likely that some intermediate cobalt concentrations (such as 28\%) have significant stacking disorder.

The change from a rhombohedral to a primitive unit cell is not especially surprising because stacking faults associated with the primitive stacking are observed in parent \FGTx crystals.\cite{Stahl2018}  Indeed, these have been associated with a structural transition at $\approx$570\,K.\cite{May2019}  Cobalt substitution appears to enhance the stability of the stacking sequence that the parent phase is unable to coherently transform to upon cooling below $\approx$570\,K, perhaps due to phase competition associated with local order/disorder.  Interestingly, we previously observed a $\sqrt{3}a\times\sqrt{3}a$ supercell in the parent Fe$_{4.87(2)}$GeTe$_2$ crystals and this was also detected in the diffraction data for the 28\% cobalt crystal. This in-plane supercell was not evident in the diffraction data for the 46\% cobalt crystal.  Due to similarity in atomic number of Fe and Co, the site occupations and total composition were not able to be refined reliably from the x-ray diffraction data.  Regardless of the approaches used, such as site preferences, the refined composition is very close to (Fe,Co)$_5$GeTe$_2$ and thus cobalt incorporation may lead to a small increase in total transition metal content.

The preferred site occupancy for cobalt was investigated with DFT calculations. Substitution of cobalt on the Fe1 sublattice is 175\,meV/Co more favorable than occupation on the Fe2 site, which is still 140\,meV/Co more favorable than occupation on the Fe3 sublattice (for the distribution of atoms on both Fe1a,b in a checkerboard arrangement with 10\% cobalt substitution).  We therefore anticipate that cobalt first substitutes (primarily) for Fe1 and then for Fe2, and have illustrated this calculation scheme in Fig.\,\ref{Structure}(b) where all Fe1 sites are occupied for simplicity.

\subsection{Evolution of Magnetic Properties}

\begin{figure}[ht!]%
\includegraphics[width=\columnwidth]{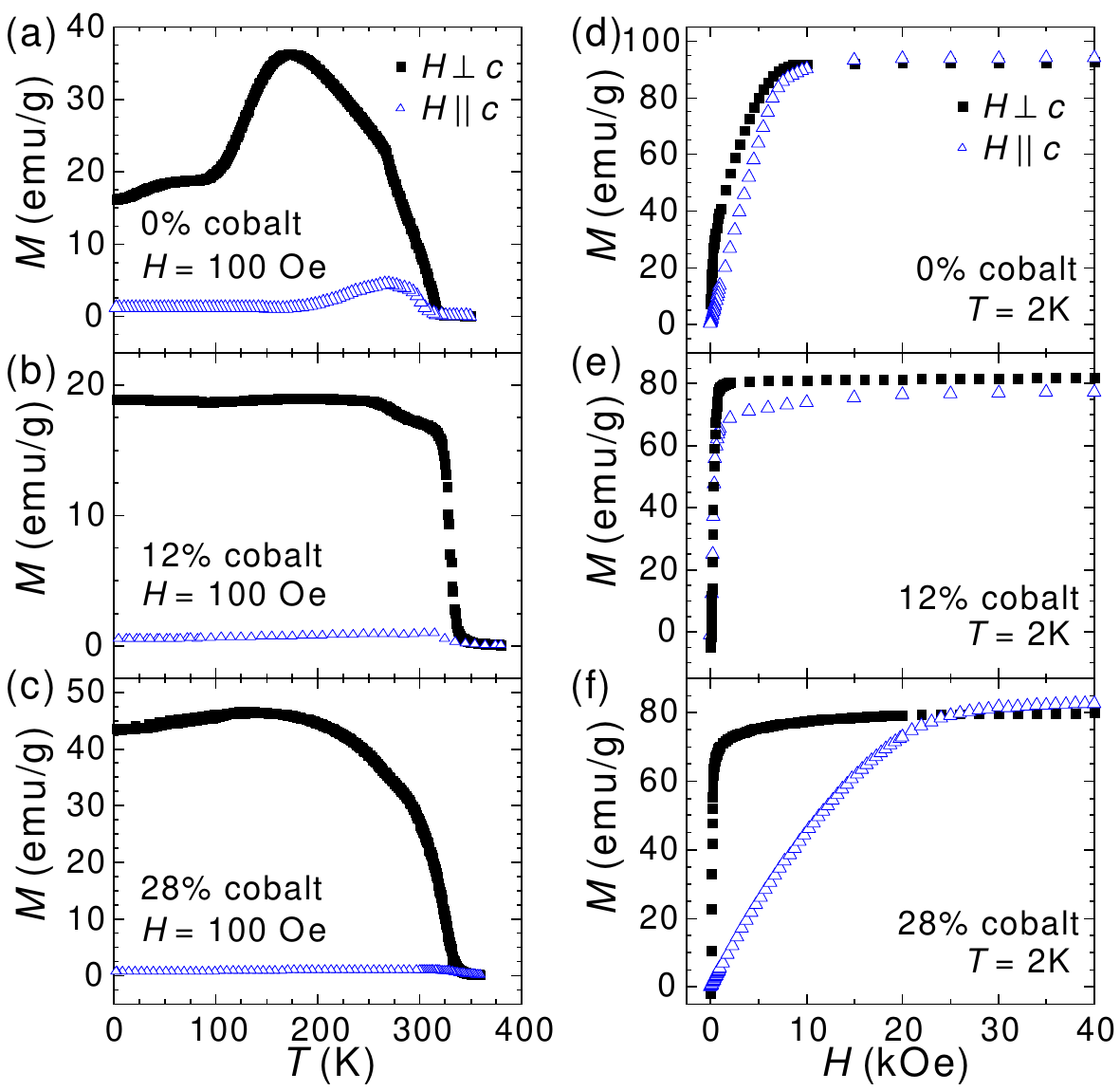}%
\caption{(Anisotropic $M$ for parent Fe$_{4.87(2)}$GeTe$_2$ in (a,d) and for the ferromagnetic cobalt-containing samples in (b,c,e,f).  The data demonstrate that cobalt substitution for iron results in a slight enhancement in the Curie temperature, along with a change in magnetic anisotropy that is best observed in (d,e,f).}%
\label{MT1}%
\end{figure}

The anisotropic magnetization data are shown in Fig.\,\ref{MT1} for crystals containing 0, 12 and 28\% cobalt. The temperature-dependent magnetization \MT in Fig\,\ref{MT1}(a-c) reveals a dominant ferromagnetic character for 12 and 28\% cobalt.  The cobalt-containing samples consistently have higher $T_{\rm C}$ than the parent phase, with \TC between 323 and 328\,K observed for 12-28\% cobalt. This \TC was determined by the intersection of linear fits  from just above and below \TC to the observed $M^2$($T$).  The increased \TC could be caused by a small change in the total transition metal content or the reduced in-plane lattice parameter upon cobalt incorporation, though previous experiments have shown that strain also appears to impact $T_{\rm C}$.\cite{ACSNano}  Due to demagnetization effects that are temperature dependent, the \MT data do not provide a clear picture of the magnetic anisotropy (the internal field becomes much smaller for \HparC upon cooling).

The isothermal magnetization data at $T$=2\,K in Fig\,\ref{MT1}(d-f) reveal a significant change in the magnetic anisotropy upon cobalt substitution for these ferromagnetic compositions.  Cobalt incorporation reduces the easy-axis anisotropy in favor of easy-plane anisotropy.  The highest cobalt content (28\%) has an easy-plane anisotropy of about 2\,T, which is roughly double that of the easy-axis anisotropy for the parent Fe$_{4.87(2)}$GeTe$_2$ crystals. This evolution of the magnetic anisotropy is illustrated in Fig.\,\ref{Structure}(b). While the trend towards easy-plane anisotropy is not necessarily desired from the perspective of stabilizing magnetic order in the pure 2D limit, controlling magnetic anisotropy via chemical substitution is an important step in tuning cleavable magnetic materials.

In addition to changing the anisotropy and enhancing $T_{\rm C}$, the incorporation of cobalt leads to a more monotonic behavior in $M$($T$).  In Fe$_{4.87(2)}$GeTe$_2$ crystals, the Fe1 sublattice is magnetically dynamic above 120\,K, and this characteristic of the magnetism dominates the low-field magnetic response as well as the transport properties.\cite{May2019}  The lack of strong anomalies below \TC in the cobalt-containing samples suggests cobalt substitution impacts the magnetic-independence of the Fe1 sublattice, likely through the preferential substitition of cobalt for Fe1 as predicted by DFT calculations.  The change in behavior of \MT with cobalt substitution also correlates with the observed change in magnetic anisotropy.  The magnetic anisotropy in the parent phase appears driven by the Fe1 moments, with the easy-axis anisotropy developing mostly below 100\,K where those moments are not dynamic.\cite{May2019} Therefore, the strong impact of cobalt substitution on the anisotropy may also support preferential occupation on the Fe1 sublattice. 

\begin{figure}[ht!]%
\includegraphics[width=\columnwidth]{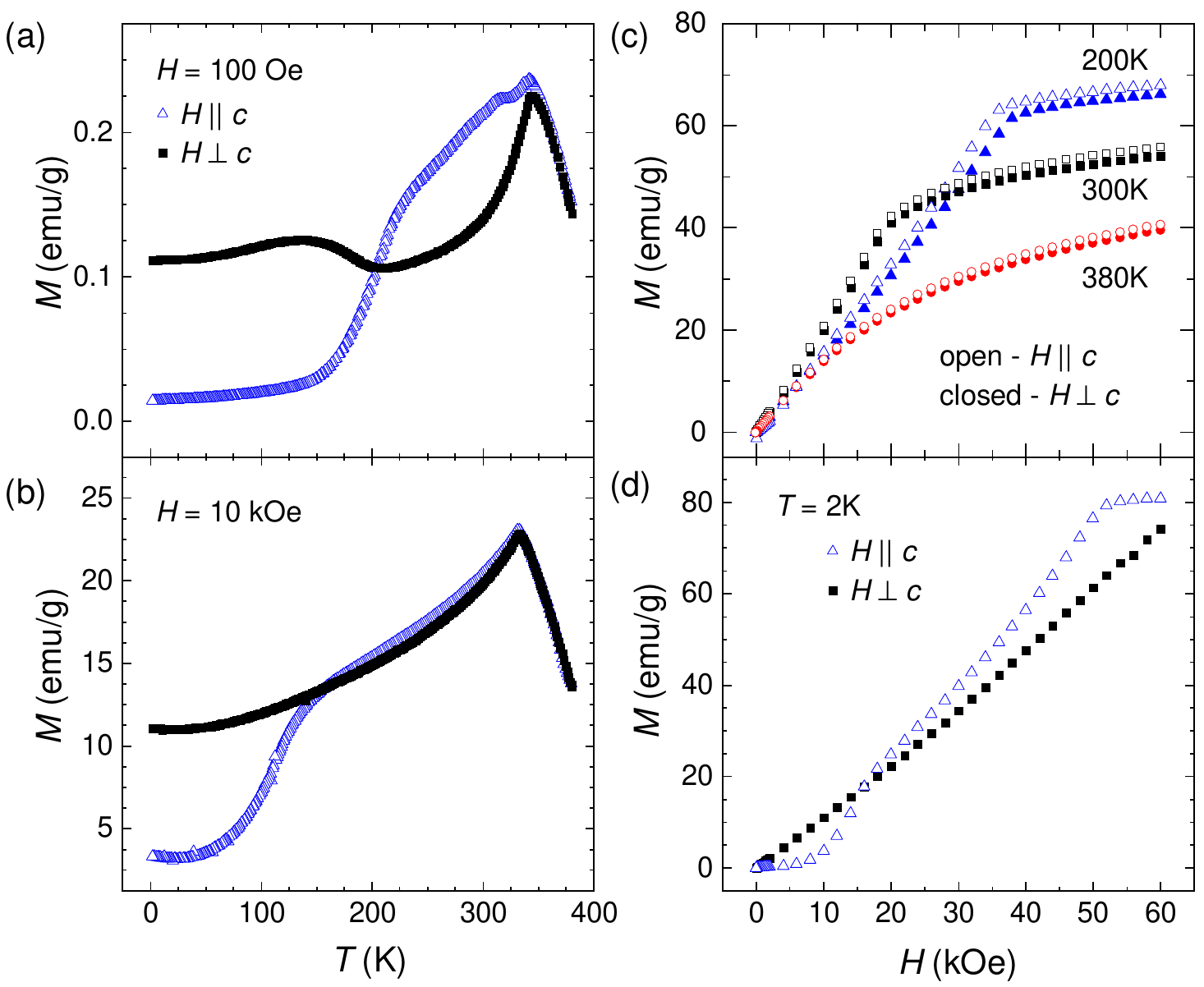}%
\caption{Magnetization data for a crystal with 46\% cobalt.  (a,b) Temperature-dependent $M$ demonstrating cusp associated with AFM order. (c,d) Isothermal magnetization data showing evolution of anisotropy with temperature. In (d) a spin-flop transition is observed near 1.5\,T at 2\,K for \HparC, which demonstrates that the moments align predominantly along [001] in zero field.}%
\label{AFM1}%
\end{figure}

In crystals with the highest cobalt concentrations (46-54\%), antiferromagnetic character was observed in the magnetization data.  Magnetization data for a crystal with 46\% cobalt content  are shown in Fig.\,\ref{AFM1}.  It is apparent that the induced magnetization is much smaller than for the ferromagnetic crystals with lower cobalt content (compare to Fig.\,\ref{MT1}(b,c)).   When measured by the cusp in $M$($T$), the N\'{e}el temperature is \TN = 343\,K ($H$=100 Oe).  This value decreases with increasing applied field, consistent with compensated AFM character.  The \MT data indicate that the orientation of the moments changes as a function of temperature.  At 200\,K and above, there is little anisotropy in the magnetic response (see Fig.\,\ref{AFM1}(c)).    At the lower temperatures, below $\approx$150\,K, the moments seemingly orient along the $c$-axis. This is evidenced in Fig.\,\ref{AFM1}(d) by the spin-flop transition near 1.5\,T for \HparC at $T$=2\,K.  Interestingly, the ordering temperature remains quite high and the saturated moment is only moderately reduced relative to that of the ferromagnetic compositions. At the highest cobalt concentrations synthesized ($\approx$54\%), dominant AFM character is observed with cusps in \MT between $\approx$290 and 310\,K.  The evolution of the anisotropy at high cobalt concentrations appears to be somewhat complex and the magnetic structure may be complicated.  We also note that the crystals remain metallic even at high cobalt concentrations (see Supplemental Materials).  Based on the calculations discussed below, we believe the AFM state contains ferromagnetic layers that are coupled antiferromagnetically along [001], and could thus be of interest for building certain types of ferromagnetic vdW heterostructures.

\begin{figure}[ht!]%
\includegraphics[width=\columnwidth]{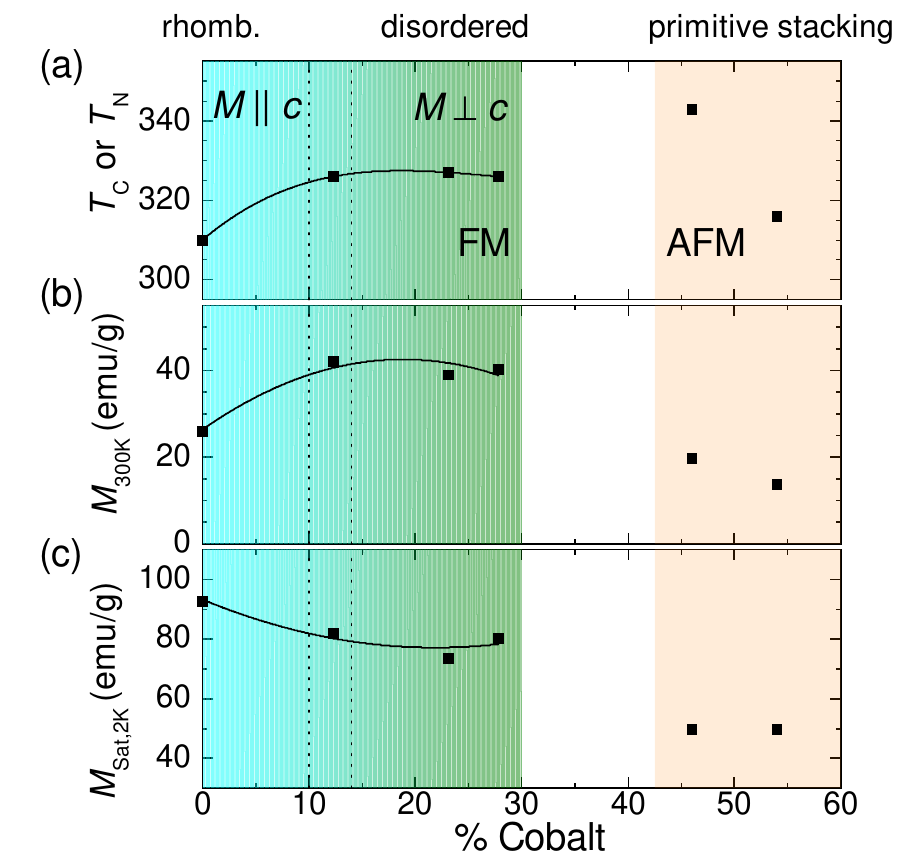}%
\caption{Summary of magnetic behavior as a function of cobalt content with layer stacking indicated along the upper horizontal axis.  (a) Curie and N\'{e}el temperatures, (b) magnetization induced along [001] at 300\,K and (c) saturation magnetization (2\,K, $H \parallel c$\,=\,60\,kOe).  The dashed vertical lines indicate the approximate region where the magnetic anisotropy inverts for the ferromagnetic compositions; the solid lines in the FM region are to facilitate viewing. The AFM region is likely characterized by FM planes that are coupled antiferromagnetically along [001].}%
\label{PD}%
\end{figure}

The impact of cobalt substitution on the magnetic response is summarized in Fig.\,\ref{PD}. Also, the schematic in Fig.\,\ref{Structure}(b) displays a change in magnetic anisotropy within the ferromagnetic portion of the phase diagram, as well as a change in interlayer coupling from FM to AFM at high cobalt content where the primitive unit cell is observed.  Figure\,\ref{PD}(a) displays the observed ordering temperatures (\TC or $T_{\rm N}$), while Fig.\,\ref{PD}(b,c) show the magnetization $M$ induced at 300\,K ($H \perp c$ = 10\,kOe) and the effective saturation magnetization at $T$=2\,K ($H$ = 60\,kOe).  The dashed lines in Fig.\,\ref{PD} enclose the approximate region where the anisotropy of the induced magnetization inverts from an easy-axis (parallel to the $c$-axis) to an easy-plane for the ferromagnetic compositions.  The  anisotropy was inferred using the disparity of critical fields for saturating $M$ in isothermal magnetization data; relatively little anisotropy is observed in the parent Fe$_{5-x}$GeTe$_2$, particularly above 100\,K, and consideration of demagnetization effects are necessary to infer the induced anisotropy (see Supplemental Materials).\cite{ACSNano}

The change from ferromagnetic to antiferromagnetic order appears to be coupled to the structural transition involving layer stacking.  The antiferromagnetic behavior in Fig.\,\ref{AFM1} and the crystal structure with a primitive unit cell in Fig.\,\ref{Structure}(b) were both obtained from crystals selected from the same crystal growth.  A similar transition between ferromagnetic order and antiferromagnetic order with layer stacking is seen in CrI$_3$, where exfoliated samples maintain the high-temperature monoclinic stacking and AFM order results. This is opposite to the FM behavior seen in bulk CrI$_3$ that has rhombohedral lattice centering at low temperature.\cite{Sivadas2018,Jiang2019,Song2019,Li2019}   

We utilized first principles calculations to probe the connection between structure, composition and magnetic order in Fe$_{5}$GeTe$_2$.  We begin by discussing results for the parent Fe$_{5}$GeTe$_2$ and then move onto the impact of cobalt substitution.  These results indicate that the intraplane coupling is predominantly ferromagnetic, but the interplane coupling is sensitive to the local atomic configurations.

The numerical results of first principles calculations are summarized in Table\,\ref{DFT} for the lowest energy configurations. For the structural configuration with Fe1a,b occupied in a checkerboard arrangement, the ferromagnetic configuration of \FGT was more stable than an antiferromagnetic configuration with AFM coupling between slabs (along [001]).    This result held for both the rhombohedral and primitive models.   The FM ground state was 10\,meV/f.u. more stable than when the slabs are coupled AFM along [001] for the rhombohedral stacking, with calculated moments of the 1.97\,$\mu_B$/Fe1, 2.04\,$\mu_B$/Fe2 and 2.42\,$\mu_B$/Fe3.  These values are similar to the $\approx$2\,$\mu_B$/Fe observed experimentally.\cite{Stahl2018,ACSNano,May2019} Calculations were also performed in the atomic configuration where the Fe1 atoms are only on the Fe1a sites. For this atomic distribution, the ferromagnetic simulation would not converge and instead a ferrimagnetic ground state was observed. This state was characterized by small Fe1 moments (-0.13$\mu_B$) coupled antiparallel to those on Fe2 and Fe3 with FM coupling between layers along [001].  However, imposing AFM coupling between the layers along [001] yielded ferromagnetic slabs with small Fe1 moments (0.12$\mu_B$) ferromagnetically aligned with the neighboring Fe2 and Fe3 moments (ferromagnetic slabs stacked antiferromagnetically).  Importantly, this AFM configuration is calculated to be 4\,meV/f.u. more stable than the ferrimagnetic state for this atomic distribution with all Fe1a occupied and Fe1b vacant.  These results thus suggest that the intraplanar coupling is dominantly ferromagnetic.  Indeed, antiferromagnetic coupling within a slab was found to be very high in energy relative to the ferromagnetic ground state, as discussed in the Supplemental Materials.

\begin{table*}[ht!]
\caption{\textbf{Energetics of local atomic order and magnetic couplings in Fe$_{5-y}$Co$_y$GeTe$_2$}. Calculated energies relative to predicted ground state and moments on the Fe1 sublattice at a given composition. The atomic distributions on the Fe1 sublattice and the interlayer coupling are modified, along with cobalt content, for rhombohedral $R$ or primitive $P$ lattices as indicated.  Positions for cobalt substitutions are illustrated in Fig.\,\ref{Structure}(b).}
\begin{tabular}[c]{|c|c|c|c|c|}
\hline
\% cobalt / lattice  & Fe1 Atomic Distribution  & FM/AFM & $\Delta$E & Moment\\
at.\% = $y/5$ &  & along [001] & meV/f.u. & $\mu_B$/Fe1 \\
\hline
\hline
0\% / $R$ & Fe1a,b checker  & FM  & 0 & 1.97 \\
0\% / $R$ & Fe1a,b stripe   & FM & 3 & 1.96\\ 
0\% / $R$ & Fe1a,b checker & AFM & 10 & 1.99\\ 
0\% / $R$ & Fe1a only & AFM  & 119 & 0.12 \\
0\% / $R$ & Fe1a only & FM & 123 & -0.13 \\ 
\hline
20\% / $R$ & Fe1a,b checker & FM & 0 & 0.68 \\
20\% / $R$ & Fe1a,b checker & AFM & 8 & 1.99\\
\hline
60\% / $P$ & Fe1a,b checker & AFM & 0 & 1.05 \\
60\% / $P$ & Fe1a,b checker & FM & 5 & 1.05\\
\hline
60\% / $R$ & Fe1a,b checker & FM & 0 & 1.05 \\
60\% / $R$ & Fe1a,b checker & AFM & 1 & 1.05\\
\hline
\end{tabular}
\label{DFT}
\end{table*}

The DFT calculations demonstrate that structural disorder and short-range atomic order will be significant in determining the magnetic properties of Fe$_{5}$GeTe$_2$.  In materials with regions of short-range atomic order that have different magnetic ground states, the neighboring domains can interact and cause a strong competition between magnetic states in a way that leads to complex magnetic behavior.\cite{Li2016}   Such an effect was observed in TlFe$_{1.6}$Se$_{2}$, where a complete spin reorientation is observed in crystals with ordering of Fe vacancies but a competition for moment orientation is observed when nanoscale phase separation exists between atomically ordered and disordered regions.\cite{May2012}  Both \FGTx and TlFe$_{1.6}$Se$_2$ possess magnetoelastic coupling evidenced by a large lattice response across a magnetic transition, which may enhance interactions between nano-domains.  Such behavior may produce the complex \MT seen in Fe$_{5-x}$GeTe$_2$, though evolving anisotropy and magnetism due to dynamic Fe1 moments are likely important as well.

The effect of cobalt substitution on the magnetism was investigated theoretically using the energetically-favorable atomic configurations (Fe1a,b site preference followed by Fe2 occupation).  At 20\% cobalt in the parent $R\bar{3}m$ structure, FM ordering remains as the ground state, albeit with a slightly reduced moment.  For DFT calculations of 60\% Co substitution (Fe$_2$Co$_3$GeTe$_2$) we first utilized the primitive unit cell obtained experimentally for a crystal with approximately 46\% cobalt content.  In this configuration, AFM coupling along [001] is more stable than the FM order by 5 meV/f.u., in agreement with the experimental result that shows AFM ordering for 45-54\% cobalt content.  Within the parent rhombohedral structure, however, FM ordering is calculated to be slightly more stable than the AFM order for 60\% cobalt (by only 1\,meV/f.u.).  It was also verified that the small increase in layer spacing that occurs with cobalt doping does not impact these qualitative results.  These results demonstrate that the transformation from FM to AFM order is correlated with the change of layer stacking and the net cobalt content.

\section{Conclusions}

This investigation has shown that the magnetism in \FGT is highly sensitive to the local atomic arrangements in a given \FGT layer as well as the interlayer stacking configuration.  Cobalt substitution allows one to control the magnetic response in \FGT at a surprising level by impacting the nature of the magnetic interactions and the atomic configurations.  Cobalt substitutions up to $\approx$30\% enhance the ferromagnetic character by substituting primarily on the Fe1 sublattice.  The result is a unified response of the different magnetic sublattices, an enhancement in \TC and an inversion of the magnetic anisotropy.  Cobalt concentrations of 45-55\% induce an antiferromagnetic ground state with relatively little magnetic anisotropy.  A transition to a primitive unit cell is also observed at similar compositions.  DFT calculations reveal that both the primitive unit cell and a high cobalt content are required to induced the AFM ground state.  In general, the calculations reveal that local atomic configurations and layer stacking strongly impact the predicted magnetic ground state, while the intralayer coupling is dominantly ferromagnetic.   As such, these results highlight how metallic, cleavable materials can be tuned to access different states, in large part because of their weak interlayer interactions.  From a chemical perspective, controlling the composition and identifying alternative synthesis strategies that can promote particular types of local order are promising avenues to further tune the magnetism in \FGT and similar itinerant materials containing structural disorder.

After the initial submission of this manuscript, an article reporting similar changes in the magnetization of \FGTx upon cobalt substitution was published.\cite{Tian2020}

\section{Outlook}

The introduction of magnetic materials into the catalogue of exfoliable vdW crystals has allowed for fundamental investigations of magnetism in the 2D limit as well as the development of devices with advanced functionality.\cite{Duong2017,McGuire-halides2017,Burch2018,Gibertini2019rev,Gong2019rev,Wang2020rev,Zhang2019rev,Huang2020rev,Mak2019rev}  A key challenge to advance these pursuits is the ability to control the critical temperature and the nature of the magnetic ground state.  The coupling of magnetic order to layer stacking was recently demonstrated in insulating CrI$_3$.\cite{Sivadas2018,Jiang2019,Song2019,Li2019}   Those impressive studies answered many of the questions that lingered from the original work on exfoliated CrI$_3$ and have highlighted the important role of layer stacking in materials where weak interlayer exchanges dominant the magnetic ordering.  In this work, a similar effect has been demonstrated in a metallic material where a simultaneous change in layer stacking and magnetic order is induced by chemical manipulation.  The general implication is that the magnetism in metallic vdW materials can be strongly sensitive to the layer stacking.  Thus, it is important for researchers to monitor the crystal structure of layered vdW materials as a function of temperature, chemical substitutions, pressure and exfoliation, and to minimize assumptions about the structure when interpreting physical behaviors.  This presents both a challenge and an opportunity for experimentalists, and with proper theoretical support could provide rational tuning and control over magnetism in these materials. 

\section{Acknowledgments}

We are pleased to thank J. Yan, B. Sales, S. Okamoto  and D. Parker for useful discussions and R. Custelcean for assistance with x-ray single-crystal diffraction measurements. This work was supported by the U. S. Department of Energy, Office of Science, Basic Energy Sciences, Materials Sciences and Engineering Division.  ORNL is managed by UT-Battelle, LLC, under Contract No. DE-AC05-00OR22725 for the U.S. Department of Energy.

The United States Government retains and the publisher, by accepting the article for publication, acknowledges that the United States Government retains a nonexclusive, paid-up, irrevocable, world-wide license to publish or reproduce the published form of this manuscript, or allow others to do so, for U. S. Government purposes. The Department of Energy will provide public access to these results of federally sponsored research in accordance with the DOE Public Access Plan.

\newpage

\onecolumngrid
\appendix

\renewcommand{\theequation}{S\arabic{equation}}
\renewcommand{\thefigure}{S\arabic{figure}}
\renewcommand{\thetable}{S\arabic{table}}

\setcounter{figure}{0}
\newpage
\section*{Supplemental Materials}

\section*{X-ray Diffraction}

Single crystal x-ray diffraction data for the crystal selected from the growth that produced 46\% cobalt crystals were found to contain sharp spots without notable streaking along $l$ (negligible stacking disorder).  A structural model was successfully generated for this sample.  The model is in space group $P\bar{3}m1$ and contains a split site `Fe1' and Ge positions in analogy to the structural model we have previously utilized for \FGTx.  The primitive structure can also be described in space group $P\bar{3}$ with essentially identical atomic positions and slightly better refinement indicators.  We note that a twin law was necessary and several different twin laws were sufficient, including a two-fold rotation about [001] or a mirror plane perpendicular to the $c$-axis. 

Refinement using all iron in the unit cell or a mixture of iron/cobalt at 50/50 yielded similar refinement indicators. Results for the 50/50 mixture of Fe/Co across all sites are shown to emphasize the importance of the substitution.  Based on DFT calculations, site preferences are expected. 

The data were collected at 220\,K for the 46\% cobalt sample and are compared to data collected at 220\,K for the quenched \FGTx sample.  The data suggest a 0.2\% increase in van der Waals gap relative to parent \FGTx while the majority of the interlayer spacing increase comes from an increase in the slab thickness (0.7\% expansion).  The basal plane lattice parameter $a$ decreases by 0.6\% in the cobalt doped sample. 

The single crystal x-ray diffraction data obtained from a crystal in the batch that yielded large crystals with 28\% cobalt by EDS were found to contain evidence for a large degree of stacking disorder.  This is similar to \FGTx crystals that are allowed to cool naturally in the furnace but with more stacking disorder for these 28\% cobalt containing samples.\cite{Stahl2018,ACSNano}  This 28\% cobalt sample also had weak superlattice reflections similar to those in the parent phase.\cite{ACSNano}  The dominance of stacking faults was also observed in quenched \FGTx crystals that were thermally-cycled to 50\,K, which is below the first-order structural transition that occurs in metastable (quenched) crystals upon cooling below 100\,K for the first time.  Note that the cobalt containing crystals were quenched from the growth temperature of 750\degrees\,C.

\begin{figure}[h!]%
\includegraphics[width=0.8\columnwidth]{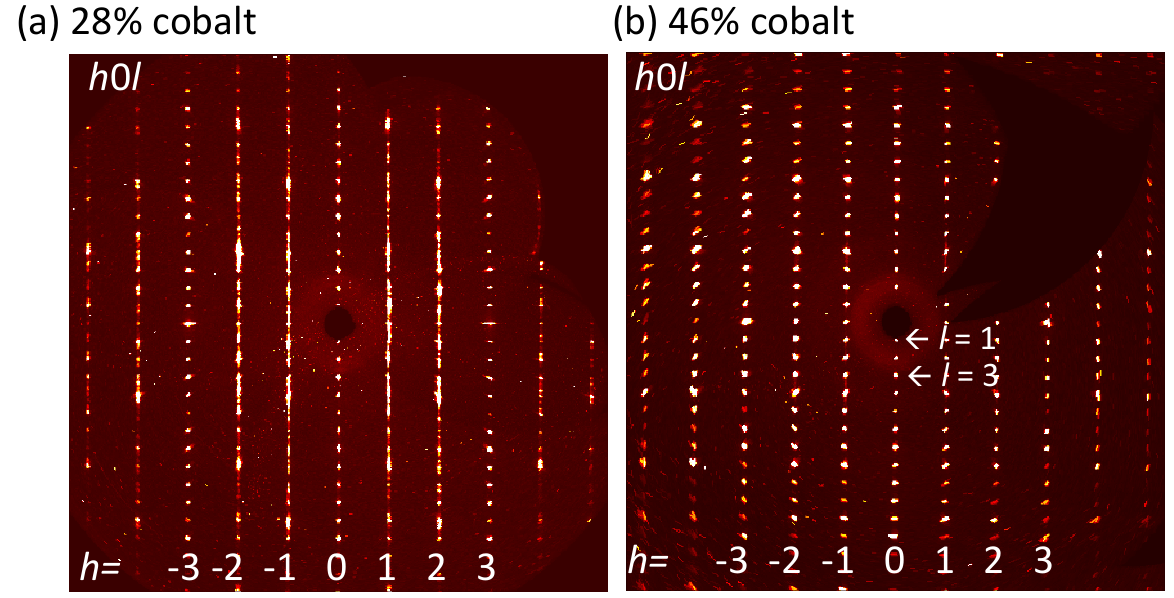}%
\caption{X-ray single crystal diffraction data viewed in the ($h0l$) plane for (a) 28\% and (b) 46\% cobalt concentration relative to iron.  The streaking along $l$ for $h$=$\pm$1 and $\pm$2 in (a) is likely dominated by mixtures of primitive and rhombohedral stacking.}%
\label{EDS}
\end{figure}


\section*{First Principles Calculations}

Structural files (.cif formats) wih relaxed atomic positions have been uploaded as additional supporting information. These files contain the atomic positions for the supercells utilized to investigate the distribution of atoms on the Fe1 sublattice.  Files are included for the atomic positions after relaxing in the ferromagnetic configuration for Fe$_5$GeTe$_2$.  The energetically preferred distribution on both Fe1a and Fe1b (a checkerboard pattern) is provided. Another distribution on Fe1a and Fe1b (stripe pattern, not included) was only 3\,meV/f.u. higher in energy and thus we expect significant disorder in real crystals with both Fe1a and Fe1b occupied.  The configuration involving only Fe1a (all-up) is significantly higher in energy (file included).  A structural file for Fe$_2$Co$_3$GeTe$_2$ is included as well, and this contains atomic positions after relaxation in the AFM ground state; two layers are necessarily included in the atomic list but the underlying symmetry is a primitive cell.

For 20\% cobalt substitution (complete substitution for Fe1 atoms), the atomic configurations with checkerboard and stripe occupations of the Fe1a,b sublattice were equal in energy.  These configurations were 63\,meV/f.u. more energetically favorable than the occupation of only Fe1a positions.  This value is for a FM state using a primitive cell to reduce computation time.

As discussed in the main text, the magnetic coupling within each layer or slab of Fe$_5$GeTe$_2$ appears to be dominantly ferromagnetic based on the calculations performed.  We also investigated a magnetic configuration with intralayer AFM coupling.  This simulation started with an AFM configuration where each crystallographic position at a given $z$-coordinate was coupled FM to itself and AFM to the neighboring sites (which are at different $z$-coordinates as moving upward through the layer). Upon relaxation, the following intra-layer AFM configuration was realized with an energy that is 438\,meV/f.u. above the calculated FM ground state.  Each atomic site listed on a separate line is at a slightly different $z$-coordinate within a given Fe$_5$GeTe$_2$ layer.  The Fe1a,b entries have fewer spins because these atoms are distributed across Fe1 in a checkerboard pattern. \\

Configuration stabilized after relaxation with AFM coupling within a given structural layer (slab) of Fe$_5$GeTe$_2$:
\begin{tabbing}
\noindent Fe1a~~~~ \= $\uparrow$ $\uparrow$ \\
\noindent Fe3 \> $\downarrow$ $\downarrow$ $\downarrow$ $\downarrow$ \\
\noindent Fe2  \> $\uparrow$ $\downarrow$ $\uparrow$ $\downarrow$ \\
\noindent Fe2  \> $\uparrow$ $\downarrow$ $\uparrow$ $\downarrow$  \\
\noindent Fe3  \> $\uparrow$ $\uparrow$ $\uparrow$ $\uparrow$ \\
\noindent Fe1b \> $\downarrow$ $\downarrow$ \\
\end{tabbing}

\begin{figure}[h!]%
\includegraphics[width=0.9\columnwidth]{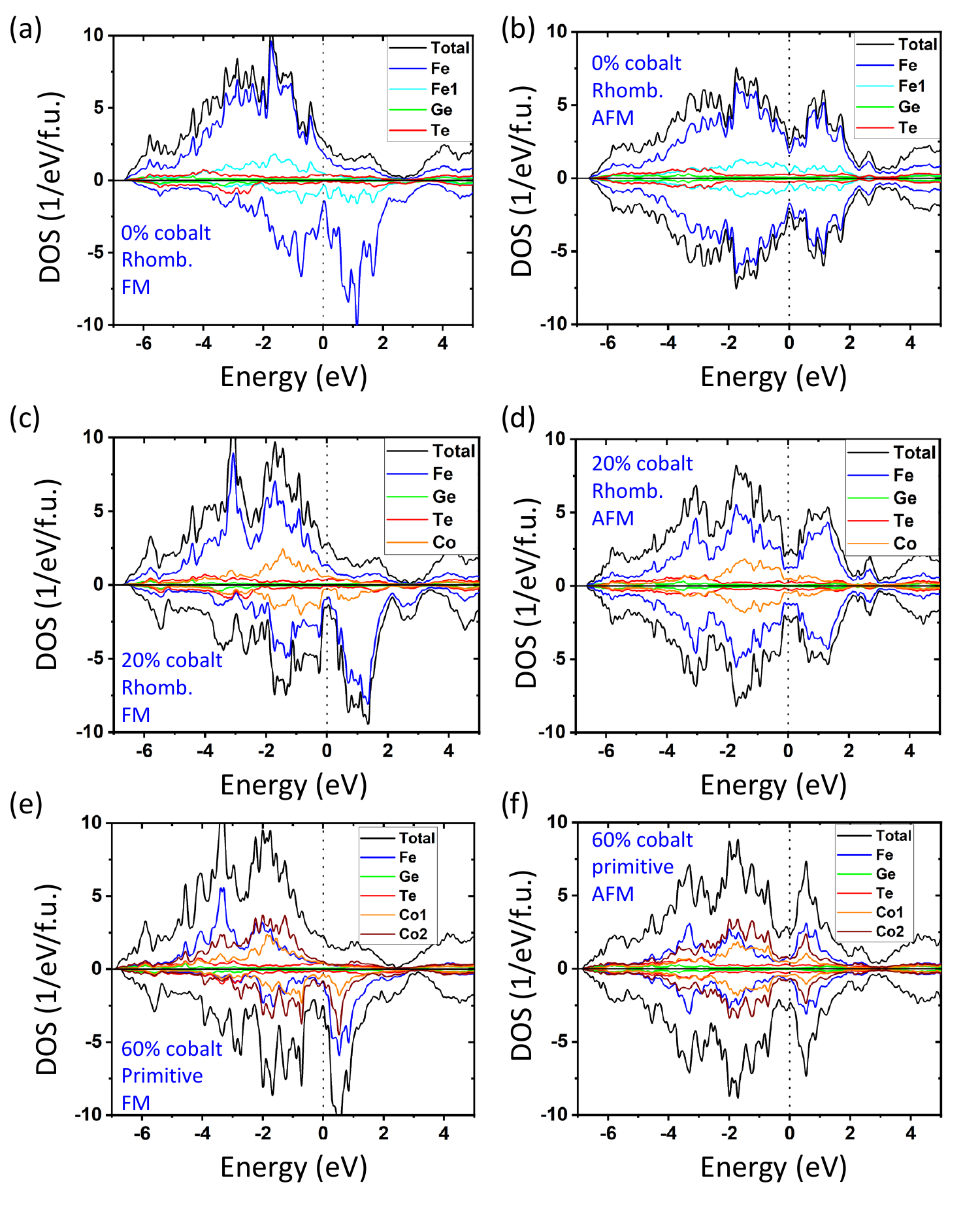}%
\caption{\textbf{ Electronic density of states (DOS) from spin polarized first principles calculations}  DOS for the ferromagnetic (FM) order in (a,c,d) are compared to DOS for the antiferromagnetic AFM ordering in (b,d,f) where AFM coupling is between slabs along [001]. Cobalt concentrations and lattice centering are listed in legend along with magnetic coupling.  All spins are ferromagnetically aligned within a structural slab.  Panels (a,c,f) are the ground state configurations for these pairwise comparisons at a given composition.}%
\label{DOS}%
\end{figure}

The projected density of states (DOS) are shown for several atomic and magnetic configurations in Fig.\,\ref{DOS}.  These calculations used a distribution of atoms on the Fe1a and Fe1b atomic positions (in a checkerboard arrangement).  The DOS are normalized as per formula unit (f.u.) so that the primitive and rhombohedral cells can be compared more directly.  The calculations for 0 and 20\% cobalt assumed the lattice constant of the parent phase, while the 60\% cobalt concentration assumed the lattice obtained by single crystal x-ray diffraction of a crystal with $\approx$46\% cobalt.  Both sets of lattice parameters come from single crystal diffraction data collected at 220\,K.  Importantly, a calculation performed for \FGT using the layer spacing associated with the primitive cell with 46\% cobalt  (accounting for symmetry change) did not have a significant effect on the total energy. 

\newpage

\section*{Magnetic Properties}

\begin{figure}[h!]%
\includegraphics[width=0.9\columnwidth]{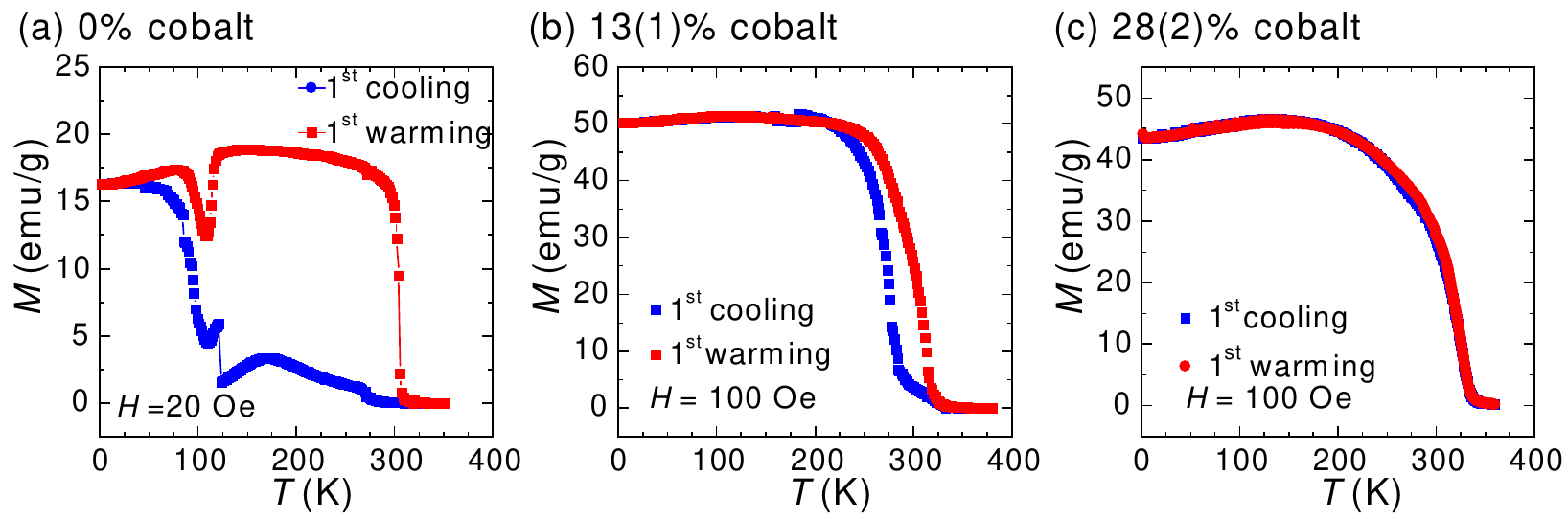}%
\caption{\textbf{Magnetization data for first thermal cycling to cryogenic temperatures after quenching from the growth conditions} Thermal hysteresis in \MT is observed in quenched crystals, as shown for (a) the parent compound and (b) a crystal with 13\% cobalt content.  This thermal hysteresis is only present during the first thermal cycling. (c) No thermal hysteresis was observed for the 28\% cobalt crystal, indicating that cobalt substitution eliminates the metastability present in Fe$_{5-x}$GeTe$_2$. Data are for \HperpC.}%
\label{Hysteresis}%
\end{figure} 

\begin{figure}[h!]%
\includegraphics[width=0.5\columnwidth]{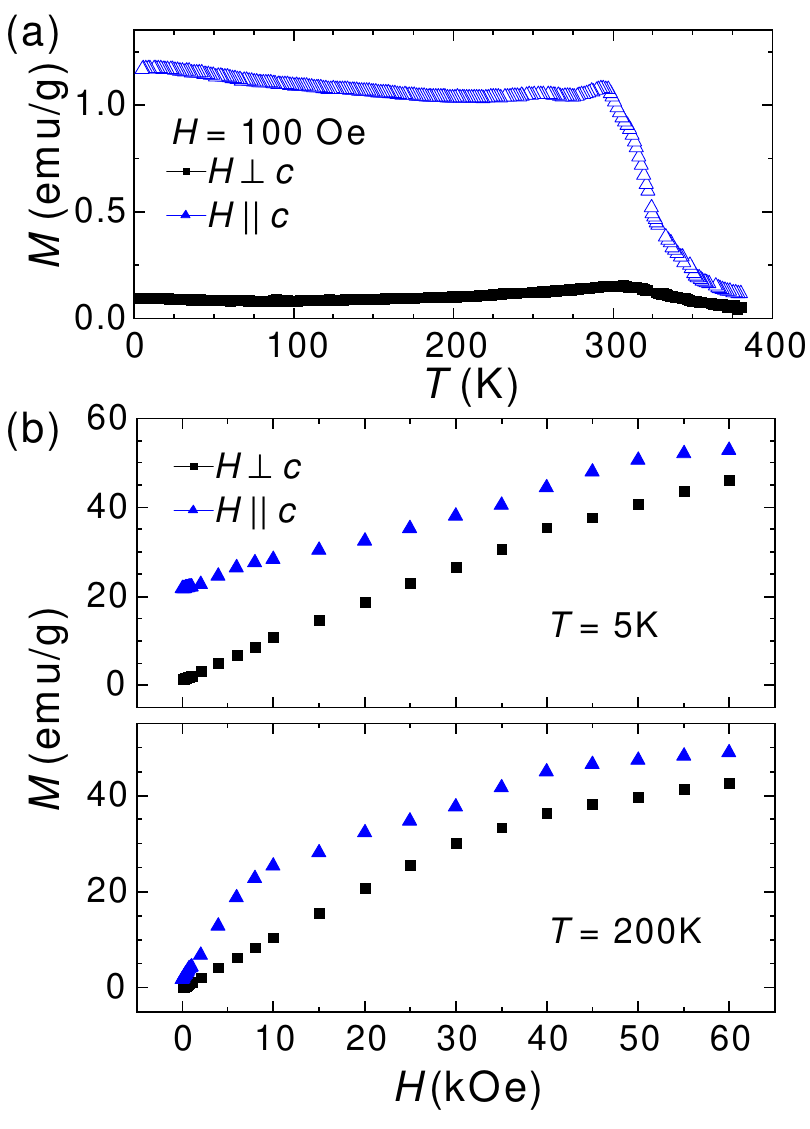}%
\caption{\textbf{Magnetization data for crystal with 54(2)\% cobalt} (a) \MT data that were collected upon cooling in an applied field; data in (b) are plotted as a function of applied field and were collected upon decreasing the magnetic field towards zero.}%
\label{172A}%
\end{figure}

\begin{figure}[h!]%
\includegraphics[width=0.9\columnwidth]{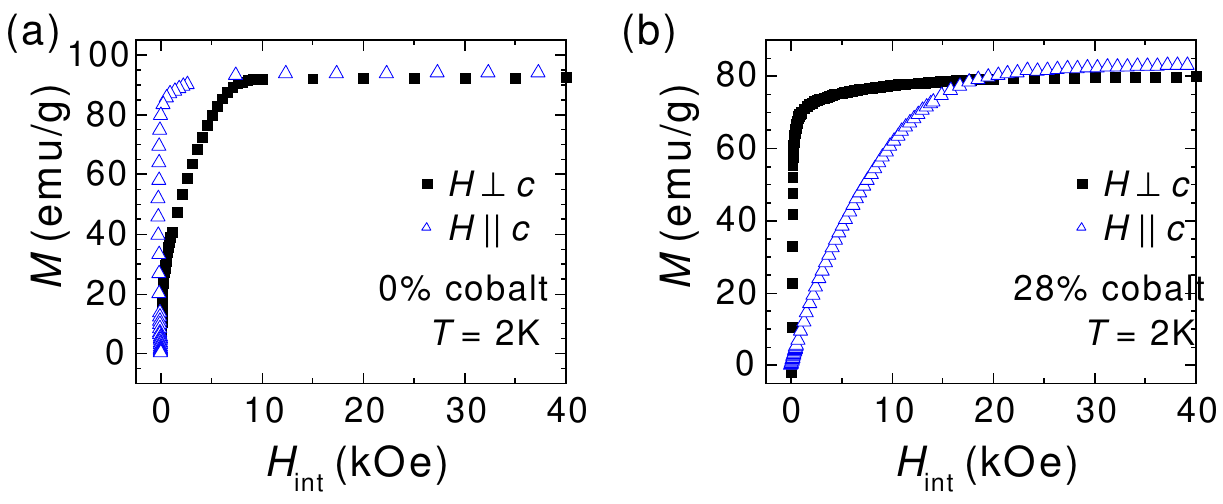}%
\caption{\textbf{Isothermal magnetization data including estimated demagnetization effects} Isothermal magnetization plotted as a function of internal field after estimating demagnetization effects for (a) \FGTx and (b) a crystal with 28\% cobalt.}%
\label{demag}%
\end{figure} 

All crystals measured here were quenched from the growth temperature. The parent \FGTx displays a first-order magneto-structural transition upon cooling below 100\,K for the first time after quenching.  This strongly impacts the magnetization and after the first thermal cycle the \TC is enhanced (to 310\,K) and the magnetic properties are reversible in the range 2-360\,K. The hysteretic \MT for the first thermal cycle of \FGTx is shown in Fig.\,\ref{Hysteresis}(a).  Hysteresis was still observed at 13\% cobalt content (Fig.\,\ref{Hysteresis}(b)) but was not observed at 28\% cobalt (Fig.\,\ref{Hysteresis}(c)).  Thermal hysteresis was not observed in the \MT data for crystals with higher cobalt content that displayed antiferromagnetic character. When thermal hysteresis was observed, the main text shows data after the first thermal cycle and thus represent the reversible properties that also possess higher \TC than upon first cooling from the quenched state. The relationship between the metastability/thermal hysteresis  and a structural transition near 570\,K in \FGTx is discussed in Ref.\,\citenum{May2019}.

Magnetization data for a crystal with a measured cobalt concentration of 54(2)\% cobalt are shown in Fig.\,\ref{172A}.  Demagnetization effects were not included.

The effect of demagnetization fields is expected to be strong for these plate-like ferromagnetic crystals.  Indeed, the \MT data would suggest that the moments are more readily aligned perpendicular to [001], yet the material appears as an easy-axis [001] ferromagnet for low $T$ according to neutron diffraction data as well as investigation of exfoliated flakes.\cite{ACSNano}  After estimating the demagnetization effect, the \MH data for \FGTx reveals that the moments are saturated with a small internal field while 7-10\,kOe is required to saturate the moments when \HperpC.  This is illustrated in Fig.\,\ref{demag}, which demonstrates that the trend with cobalt content is independent of this demagnetization effect. For simplicity, the data in the main text are plotted without estimating a demagnetization field (the fields provided are the applied magnetic field).

\section*{Specific Heat and Electrical Resistivity}

\begin{figure}[h!]%
\includegraphics[width=0.9\columnwidth]{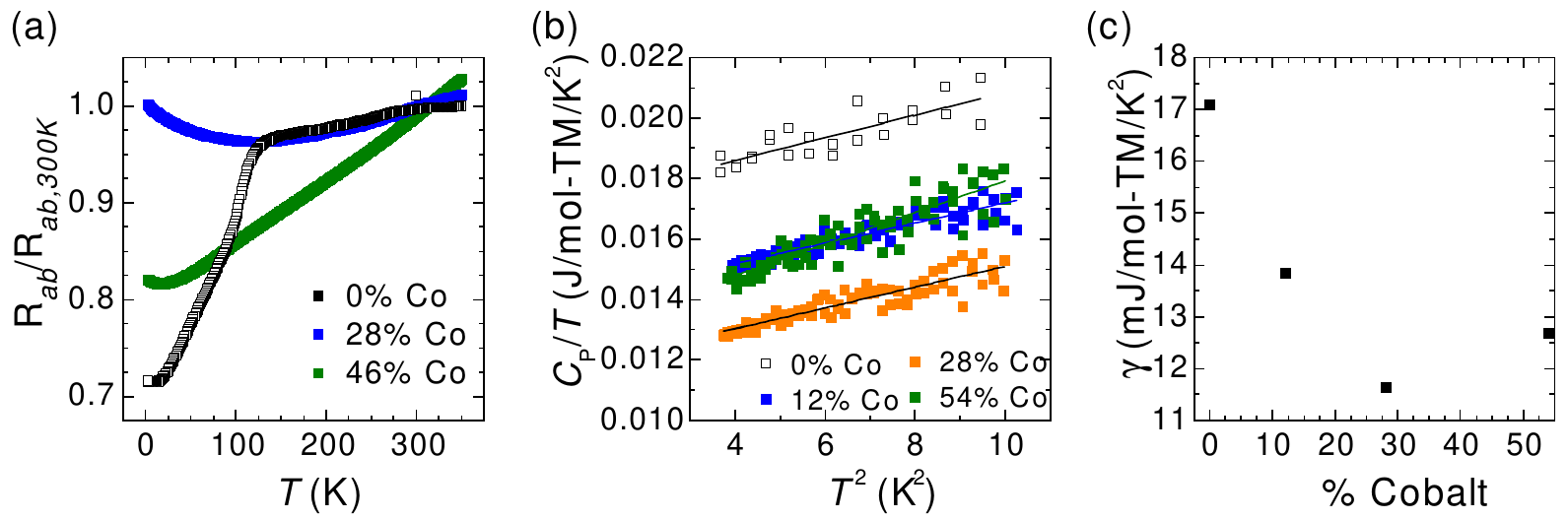}%
\caption{\textbf{Electrical resistance and specific heat capacity of select crystals}  (a) Resistance normalized to values at 300\,K. The parent phase has a strong change in temperature dependence near 120\,K associated with moments on the Fe1 sublattice and this is not observed in the cobalt containing samples. (b) Specific heat divided by temperature versus temperature squared to demonstrate the electronic contributions. (c) Sommerfeld coefficients $\gamma$ obtained from the fits in panel (b) for various cobalt concentrations.}%
\label{Resist}%
\end{figure}

Four-probe electrical resistivity and specific heat capacity measurements were completed in a Quantum Design Physical Property Measurement System to verify the metallic nature for both FM and AFM compositions.  The resistances relative to the values obtained at 300\,K are shown in Fig.\,\ref{Resist} for select crystals.   At 300\,K, the resistivity for different crystals was between 220-540$\mu$Ohm-cm.  The Sommerfeld electronic coefficient $\gamma$ was obtained from a linear fit of $C_P/T$ versus $T^2$ where the intercept is $\gamma$.  The values of $\gamma$ are per transition metal (TM) and not per formula unit.   For simplicity, the same moleclar weight is used for all compositions.  The difference in molecular weight between \FGT and hypothetical Co$_5$GeTe$_2$ is $\approx$2.5\%, which is much smaller than the change in $\gamma$ that is observed. 

\end{document}